# Fundamental governing equations of motion in consistent continuum mechanics

Ali R. Hadjesfandiari, Gary F. Dargush

*Department of Mechanical and Aerospace Engineering*
*University at Buffalo, The State University of New York, Buffalo, NY 14260 USA*

ah@buffalo.edu, gdargush@buffalo.edu

October 1, 2018

**Abstract**

We investigate the consistency of the fundamental governing equations of motion in continuum mechanics. In the first step, we examine the governing equations for a system of particles, which can be considered as the discrete analog of the continuum. Based on Newton's third law of action and reaction, there are two vectorial governing equations of motion for a system of particles, the force and moment equations. As is well known, these equations provide the governing equations of motion for infinitesimal elements of matter at each point, consisting of three force equations for translation, and three moment equations for rotation. We also examine the character of other first and second moment equations, which result in non-physical governing equations violating Newton's third law of action and reaction. Finally, we derive the consistent governing equations of motion in continuum mechanics within the framework of couple stress theory. For completeness, the original couple stress theory and its evolution toward consistent couple stress theory are presented in true tensorial forms.

# 1. Introduction

The governing equations of motion in continuum mechanics are based on the governing equations for systems of particles, in which the effect of internal forces are cancelled based on Newton's third law of action and reaction. These equations are the force and moment equations, which express, respectively, the rate of change of linear and angular momentum of the system under the influence of external forces. As every student of mechanics knows, these two vectorial governing equations are enough to describe the motion of rigid bodies (Goldstein, 1980; Meriam and Kraige, 2012). This set consists of six scalar governing equations, with three force equations corresponding to the translational motion, and three moment equations relating to the rotational motion.

In continuum mechanics, the method of analysis is based on considering the motion of infinitesimal elements of matter at each point. This means we break the continuum into infinitesimal elements at each point and then analyze the motion of these elements under the influence of stresses by using the governing equations of motion and the constitutive relations. This shows that the degrees of freedom must be the displacements and rotations at each point, which describe the translation and rotation of the infinitesimal elements of matter. Consequently, the governing equations of motion at each point in differential form are similar to those describing the motion of a rigid body.

We will show here that although defining other moments and moments of momentum for a system of particles is possible, such as symmetric first and second moments and the symmetric moment of momentum, these do not result in new physical governing equations for the system. This is because Newton's third law of action and reaction does not allow cancellation of the effect of internal forces in these moment equations.

Mindlin and Tiersten (1962) and Koiter (1964) developed the initial version of couple stress theory (MTK-CST), in which the deformation is completely specified by the continuous displacement field. This theory is implicitly based on the rigid body portion of motion of infinitesimal elements of matter at each point of the continuum (Hadjesfandiari and Dargush, 2011, 2015b). Therefore, in this theory, the internal stresses are exactly the force- and couple-stress tensors, introduced by Cosserat and Cosserat (1909), which each have at most nine independent components. However, the final MTK-CST suffers from some serious inconsistencies and difficulties with the underlying formulations (Hadjesfandiari and Dargush, 2015a,b), such as the indeterminacy of the couple-stress tensor. Eringen (1968) realized this inconsistency as a major mathematical problem in the original MTK-CST, which he afterwards called indeterminate couple stress theory.

Consequently, the first step in using couple stress theory should be answering the criticism of Eringen (1968) about the indeterminacy of the couple-stress tensor in MTK-CST. However, Mindlin and others were not able to resolve this trouble within the framework of MTK-CST. As a result, Mindlin never used MTK-CST to solve any problem after Eringen's criticism.

The literature shows that the original MTK-CST always has been fascinating. In this theory, the deformation is completely specified by the continuous displacement field, and the state of stresses are described by the physical force- and couple-stress tensors introduced by Cosserat and Cosserat (1909). By neglecting the indeterminacy, this theory predicts reasonable solutions for two-dimensional deformation of linear isotropic elastic materials. Therefore, it is natural to expect that resolving the indeterminacy issue must happen in the general framework of MTK-CST, which in turn may save continuum mechanics from so much confusion. This fortunately has happened by discovering the subtle skew-symmetric character of the couple-stress tensor (Hadjesfandiari and Dargush, 2011). Amazingly, in this consistent couple stress theory (C-CST), the couple-stress tensor is fully determinate. C-CST resolves the quest for a consistent size-dependent continuum mechanics by answering the criticism of Eringen, and provides a fundamental basis for the development of size-dependent material response.

For clarity, we also inspect some aspects of the original MTK-CST and its evolution toward consistent couple stress theory (C-CST). This includes presenting the theory by using the true third order form of the couple-stress tensor, which can be very illuminating and useful for comparing other existing higher order theories.

The remainder of the paper is organized as follows. In Section 2, we briefly investigate the character of the governing equations of motion for a system of particles in mechanics. Next, in Section 3, we demonstrate that defining new higher moments of forces and moments of momentum will not result in additional physical governing moment equations, because these violate Newton's third law of action and reaction. In Section 4, we investigate the character of the governing equations of motion for a continuum in the framework of the original couple stress theory (MTK-CST), which also includes a presentation of this theory based on true tensors. Afterward, in Section 5, we present consistent couple stress theory (C-CST) from a similar new perspective. Finally, Section 6 contains a summary and some general conclusions.

## 2. Fundamental governing equations of motion for a system of particles

Consider the motion of a particle with mass $m$ under the influence of the resultant force $F_i$. We specify the location of the particle in space at time $t$ by the position vector $x_i(t)$. The velocity and acceleration vectors are defined as $v_i = \dfrac{dx_i}{dt}$ and $a_i = \dfrac{dv_i}{dt}$, and the linear momentum is defined as $P_i = mv_i$. The governing equations of motion of the particle are given by Newton's second law as

$$F_i = \frac{dP_i}{dt} = ma_i \qquad \mathbf{F} = \frac{d\mathbf{P}}{dt} = m\mathbf{a} \qquad (1)$$

This vector equation, which is also called the force equation, states that the rate of change of linear momentum equals the resultant force acting on the particle.

The moment of the force $M_i$ and the angular momentum $L_i$ about the origin are defined, respectively, as

$$M_i = \varepsilon_{ijk} x_j F_k \qquad \mathbf{M} = \mathbf{x} \times \mathbf{F} \qquad (2)$$

$$L_i = \varepsilon_{ijk} x_j P_k \qquad \mathbf{L} = \mathbf{x} \times \mathbf{P} \qquad (3)$$

where $\varepsilon_{ijk}$ represents the permutation or Levi-Civita symbol. We notice that

$$M_i = \frac{dL_i}{dt} = \varepsilon_{ijk} x_j m a_k \qquad \mathbf{M} = \frac{d\mathbf{L}}{dt} = \mathbf{x} \times m\mathbf{a} \qquad (4)$$

This vector equation, which is also called the moment equation for the particle, states that the rate of change of angular momentum equals the resultant moment.

It should be noticed that the moment $M_i$ and the angular momentum $L_i$ are pseudo-vectors. This means these vectors can be represented by the true skew-symmetric tensor $M_{ij}$ and the angular momentum tensor $L_{ij}$, where

$$M_{ij} = \frac{1}{2}\left(F_i x_j - F_j x_i\right) \qquad (5)$$

$$\begin{aligned} L_{ij} &= \frac{1}{2}\left(P_i x_j - P_j x_i\right) \\ &= \frac{1}{2}\left(m v_i x_j - m v_j x_i\right) \end{aligned} \qquad (6)$$

Notice that $M_{ji} = -M_{ij}$ and $L_{ji} = -L_{ij}$. Interestingly, we have the dual relations

$$M_i = \varepsilon_{ijk} M_{kj} \qquad M_{ij} = \frac{1}{2}\varepsilon_{jik} M_k \qquad (7)$$

$$L_i = \varepsilon_{ijk} L_{kj} \qquad L_{ij} = \frac{1}{2}\varepsilon_{jik} L_k \qquad (8)$$

As a result, the pseudo vectorial moment equation (4) can be written in a true tensorial form

$$M_{ij} = \frac{dL_{ij}}{dt} = \frac{1}{2}\left(m a_i x_j - m a_j x_i\right) \qquad (9)$$

Although this tensorial form of the moment equation is not popular, it will be very instructive for our investigations in this paper.

Now consider a system of particles interacting with each other. For each particle, we consider the equations (1) and (4). We notice that the resultant force $F_i$ for each particle can be decomposed as

$$F_i = F_i^{ext} + F_i^{int} \qquad (10)$$

Here $F_i^{ext}$ is the external resultant force, whereas $F_i^{int}$ represents the internal resultant force on the particle exerted by other particles in the system.

By adding the force and moment equations for all individual particles in the system, we obtain

$$\sum F_i = \frac{d}{dt}\sum P_i = \sum ma_i \qquad \sum \mathbf{F} = \frac{d}{dt}\sum \mathbf{P} = \sum m\mathbf{a} \qquad (11)$$

$$\sum M_i = \frac{d}{dt}\sum L_i = \varepsilon_{ijk}\sum x_j ma_k \qquad \sum \mathbf{M} = \frac{d}{dt}\sum \mathbf{L} = \sum \mathbf{x} \times m\mathbf{a} \qquad (12)$$

where $\sum P_i$ is the total linear momentum of particles, and $\sum L_i$ is the total angular momentum about the origin. We notice that the total force $\sum F_i$ and the total moment $\sum M_i$ are a combination of the external and internal forces and moments, respectively, where

$$\sum F_i = \sum F_i^{ext} + \sum F_i^{int} \qquad (13)$$

$$\sum M_i = \sum M_i^{ext} + \sum M_i^{int} \qquad (14)$$

However, due to Newton's third law of action and reaction, the effect of internal forces and moments disappear, that is $\sum F_i^{int} = 0$ and $\sum M_i^{int} = 0$ (Goldstein, 1980; Meriam and Kraige, 2012). Therefore, the force and moment equations for the system of particles become

$$\sum F_i^{ext} = \frac{d}{dt}\sum P_i = \sum ma_i \qquad (15)$$

$$\sum M_i^{ext} = \frac{d}{dt}\sum L_i = \varepsilon_{ijk}\sum x_j ma_k \tag{16}$$

It turns out these two vectorial equations are the only possible equations for the system of particles, in which the internal forces and internal moments are cancelled based on Newton's third law of action and reaction. This is the reason why these equations are considered as fundamental governing equations of the system of particles. Although defining higher moments of forces and momentum is possible, they do not have any significance from a physical point of view. It is only for the first vectorial moment equation (16) that the moment of internal forces are cancelled, as a result of Newton's third law of action and reaction. We also notice that the first moment equation about any other arbitrary point can be written as a linear combination of force and moment equations (15) and (16).

It is obvious that the vectorial moment equation (16) can also be written as the true tensorial equation

$$\begin{aligned}\sum M_{ij}^{ext} &= \sum \frac{1}{2}\left(ma_i x_j - ma_j x_i\right) \\ &= \frac{d}{dt}\sum L_{ij}\end{aligned} \tag{17}$$

where $\sum M_{ij}^{ext}$ and $\sum L_{ij}$ are the skew-symmetric tensors, corresponding to $\sum M_i^{ext}$ and $\sum L_i$, respectively.

## 3. Inconsistency of other moment and momentum equations

For more insight, we investigate the character of some other moment and moment of momentum equations for a system of particles. We may consider the first general moment $F_i x_j$ or its symmetric part $\frac{1}{2}\left(F_i x_j + F_j x_i\right)$. We may even consider higher moments, such as $F_i x_j x_k$, $\frac{1}{2}(F_i x_j + F_j x_i) x_k$ or $\frac{1}{2}(F_i x_j - F_j x_i) x_k$, in our investigation. However, Newton's third law of action and reaction shows that these quantities have no physical significance and cannot be used to derive new consistent higher moment governing equations. For future reference, we examine some of these higher moments and corresponding angular momentum equations.

### 3.1. First general moment and angular momentum equations

We may define the first general moment tensor $\widehat{M}_{ij}$ of the force $F_i$ and the first general angular momentum tensor $\widehat{L}_{ij}$ of the particle as

$$\widehat{M}_{ij} = F_i x_j \tag{18}$$

$$\widehat{L}_{ij} = P_i x_j = m v_i x_j \tag{19}$$

Each of these generally non-symmetric true (polar) tensors is specified by nine components. We notice that

$$\frac{d\widehat{L}_{ij}}{dt} = m v_i v_j + m a_i x_j \tag{20}$$

Then, by using Newton's second law for the particle (1), we obtain

$$\widehat{M}_{ij} = m a_i x_j = \frac{d\widehat{L}_{ij}}{dt} - m v_i v_j \tag{21}$$

This is the first general moment tensor equation for the particle. However, this equation does not play any significant role in mechanics, as we demonstrate.

For the system of particles, we obtain

$$\sum \widehat{M}_{ij} = \sum m a_i x_j$$
$$= \frac{d}{dt} \sum \widehat{L}_{ij} - \sum m v_i v_j \tag{22}$$

where $\sum \widehat{M}_{ij}$ is the total first general moment of all forces, and $\sum \widehat{L}_{ij}$ is the total first general angular momentum of all particles about the origin. We notice that the total moment $\sum \widehat{M}_{ij}$ is a combination of the external moments $\sum \widehat{M}_{ij}^{ext}$ and internal moments $\sum \widehat{M}_{ij}^{int}$, where

$$\sum \widehat{M}_{ij} = \sum \widehat{M}_{ij}^{ext} + \sum \widehat{M}_{ij}^{int} \tag{23}$$

Therefore, for the system of particles, the general moment equation is written as

$$\sum \widehat{M}_{ij}^{ext} + \sum \widehat{M}_{ij}^{int} = \sum m a_i x_j$$
$$= \frac{d}{dt} \sum \widehat{L}_{ij} - \sum m v_i v_j \tag{24}$$

However, the effect of the resultant internal general moments $\sum \widehat{M}_{ij}^{int}$ does not disappear. Therefore, the left hand side of (24) includes the first general moment of internal forces. This is the reason why the first general moment equation (24), although perfectly valid, is not a fundamental governing equation and does not have any significance from a practical point of view.

### 3.2. First symmetric moment and symmetric angular momentum equations

We may define the first symmetric moment $\tilde{M}_{ij}$ of the force $F_i$ and the first symmetric angular momentum tensor $\tilde{L}_{ij}$ of the particle as

$$\tilde{M}_{ij} = \frac{1}{2}\left(F_i x_j + F_j x_i\right) = \tilde{M}_{ji} \tag{25}$$

$$\tilde{L}_{ij} = \frac{1}{2}\left(P_i x_j + P_j x_i\right)$$
$$= \frac{1}{2}\left(m v_i x_j + m v_j x_i\right) \tag{26}$$

where $\tilde{M}_{ji} = \tilde{M}_{ij}$ and $\tilde{L}_{ji} = \tilde{L}_{ij}$. Each of these symmetric true tensors is specified by six components. We notice that

$$\frac{d\tilde{L}_{ij}}{dt} = \frac{1}{2}\left(2mv_i v_j + ma_i x_j + ma_j x_i\right) \tag{27}$$

Then, by using Newton's second law for the particle (1), we obtain

$$\tilde{M}_{ij} = \frac{1}{2}\left(ma_i x_j + ma_j x_i\right) = \frac{d\tilde{L}_{ij}}{dt} - mv_i v_j \tag{28}$$

which is the first symmetric moment tensor equation for the particle. Again, this equation is perfectly valid, but does not play any significant role in mechanics, as we demonstrate.

For the system of particles, we obtain

$$\begin{aligned}\sum \tilde{M}_{ij} &= \sum \frac{1}{2}\left(ma_i x_j + ma_j x_i\right) \\ &= \frac{d}{dt}\sum \tilde{L}_{ij} - \sum mv_i v_j\end{aligned} \tag{29}$$

where $\sum \tilde{M}_{ij}$ is the total first symmetric moment of all forces, and $\sum \tilde{L}_{ij}$ is the total first symmetric angular momentum of all particles about the origin. This relation can also be written as

$$\begin{aligned}\sum \tilde{M}_{ij}^{ext} + \sum \tilde{M}_{ij}^{int} &= \sum \frac{1}{2}\left(ma_i x_j + ma_j x_i\right) \\ &= \frac{d}{dt}\sum \tilde{L}_{ij} - \sum mv_i v_j\end{aligned} \tag{30}$$

where $\sum \tilde{M}_{ij}$ is a combination of the external moments $\sum \tilde{M}_{ij}^{ext}$ and internal moments $\sum \tilde{M}_{ij}^{int}$.

However, we notice that the effect of the resultant internal symmetric moments $\sum \tilde{M}_{ij}^{int}$ does not disappear. Therefore, the left hand side of (30) includes the first symmetric moment of internal forces. This shows that the first symmetric moment equation (30) is not a fundamental governing equation of mechanics.

### 3.3. Second moment and second angular momentum equations

We may define different second moment and angular momentum tensors of the particle. Here, we only examine the second symmetric moment and angular momentum as

$$\tilde{M}_{ijk} = \tilde{M}_{jik} = \frac{1}{2}\left(F_i x_j + F_j x_i\right) x_k \tag{31}$$

$$\tilde{L}_{ijk} = \tilde{L}_{jik} = \frac{1}{2}\left(P_i x_j + P_j x_i\right) x_k = \frac{1}{2}\left(mv_i x_j + mv_j x_i\right) x_k \tag{32}$$

We notice that each of these tensors is specified by 18 components. By taking the derivative of the second symmetric angular momentum tensor (32) with respect to time, we obtain

$$\begin{aligned}\tilde{M}_{ijk} &= \frac{1}{2}\left(ma_i x_j + ma_j x_i\right) x_k \\ &= \frac{d\tilde{L}_{ijk}}{dt} - \frac{1}{2}\left(mv_i v_j + mv_j v_i\right) x_k - \frac{1}{2}\left(mv_i x_j + mv_j x_i\right) v_k\end{aligned} \tag{33}$$

This is the second symmetric moment tensor equation for the particle.

For the system of particles, we obtain the second symmetric moment equation as

$$\begin{aligned}\sum \tilde{M}_{ijk} &= \sum \frac{1}{2}\left(ma_i x_j + ma_j x_i\right) x_k \\ &= \frac{d}{dt}\sum \tilde{L}_{ijk} - \sum\left[\frac{1}{2}\left(mv_i v_j + mv_j v_i\right) x_k + \frac{1}{2}\left(mv_i x_j + mv_j x_i\right) v_k\right]\end{aligned} \tag{34}$$

where $\sum \tilde{M}_{ijk}$ is the total second symmetric moment of all forces, and $\sum \tilde{L}_{ijk}$ is the total second symmetric angular momentum of all particles about the origin. Again, we notice that the total symmetric moment $\sum \tilde{M}_{ijk}$ with 18 independent components is a combination of the external and internal second symmetric moments $\sum \tilde{M}_{ijk}^{ext}$ and $\sum \tilde{M}_{ijk}^{int}$, where

$$\sum \tilde{M}_{ijk} = \sum \tilde{M}_{ijk}^{ext} + \sum \tilde{M}_{ijk}^{int} \tag{35}$$

Therefore, for the system of particles, the second symmetric moment equation is written as

$$\sum \tilde{M}_{ijk}^{ext} + \sum \tilde{M}_{ijk}^{int} = \sum \frac{1}{2}\left(ma_i x_j + ma_j x_i\right) x_k \qquad (36)$$
$$= \frac{d}{dt}\sum \tilde{\tilde{L}}_{ijk} - \sum \left[\frac{1}{2}\left(mv_i v_j + mv_j v_i\right) x_k + \frac{1}{2}\left(mv_i x_j + mv_j x_i\right) v_k\right]$$

However, equation (36) does not define a fundamental governing equation, because the effect of the resultant internal second symmetric moments $\sum \tilde{M}_{ijk}^{int}$ does not disappear.

*3.4. Summary*

We have demonstrated that the force and moment equations

$$\sum F_i^{ext} = \sum ma_i = \frac{d}{dt}\sum P_i \qquad (37)$$

$$\sum M_i^{ext} = \varepsilon_{ijk}\sum x_j ma_k = \frac{d}{dt}\sum L_i \qquad (38)$$

are the only possible governing equations for a system of particles, in which the internal forces are cancelled based on Newton's third law of action and reaction. Therefore, these equations are the fundamental governing equations of motion for the system of particles, resulting from Newton's second and third laws. Although the pseudo (axial) vectorial moment equation (38) is always preferred in practice, we notice that it is a different representation of the true (polar) skew-symmetric moment equation (17). As we know, the force and moment equations (37) and (38) are enough to describe the motion of a rigid body (Goldstein, 1980; Meriam and Kraige, 2012), where each scalar equation for an individual component of the equations (37) and (38) describes the motion corresponding to a degree of freedom of the rigid body. In Section 4, we demonstrate in detail that the fundamental governing equations in couple stress theory (CST) are based on the force and moment equations (37) and (38).

We note that when the resultant external forces vanishes, i.e., $\sum F_i^{ext} = 0$, the linear momentum of the system is conserved. If the external moment about some point is zero, that is $\sum M_i^{ext} = 0$, the angular momentum is conserved about that point. Based on Noether's theorem (Noether, 1918), the conservation laws are the result of the symmetry properties of nature. Therefore, we notice that the conservation laws of linear and angular momentum are the result of the translational and rotational symmetry of space, respectively.

Although defining higher moments of forces and momentum is possible, they do not have any significance from a physical point of view, because the effect of internal forces do not disappear. For example, we have demonstrated that the force and the first general moment equations

$$\sum F_i^{ext} = \sum ma_i \tag{39}$$

$$\sum \widehat{M}_i^{ext} + \sum \widehat{M}_{ij}^{int} = \sum ma_i x_j \tag{40}$$

cannot describe the motion of any system, because the effect of the general moment of internal forces $\sum \widehat{M}_{ij}^{int}$ does not disappear. By taking $\sum \widehat{M}_{ij}^{int} = 0$, we obtain the inconsistent governing equations

$$\sum F_i^{ext} = \sum ma_i \tag{41}$$

$$\sum \widehat{M}_{ij}^{ext} = \sum ma_i x_j \tag{42}$$

for the system of particles. In an incorrect mechanics based on the governing equations (41) and (42), couples of forces are represented by their general moment tensors. However, we notice that the general moment tensor of a couple depends on the position of origin, and does not have any significance.

It is obvious that the 12 scalar equations (41) and (42) cannot describe the motion of a rigid body correctly. Therefore, the governing equations (41) and (42) cannot be used as a basis in continuum mechanics to describe the motion of an infinitesimal element of matter. However, it turns out that these equations are already in widespread use in size-dependent continuum mechanics with disastrous consequences.

Similarly, we have demonstrated that the force and the first symmetric moment equations

$$\sum F_i^{ext} = \sum ma_i \tag{43}$$

$$\sum \tilde{M}_{ij}^{ext} = \sum \frac{1}{2}\left(ma_i x_j + ma_j x_i\right) \tag{44}$$

cannot describe the motion of any system, because the effect of the symmetric moment of internal forces $\sum \tilde{M}_{ij}^{int}$ does not disappear. In an incorrect mechanics based on the governing equations (43) and (44), the symmetric moment tensor of a couple depends on the position of the origin, and does not have any significance. We notice that the nine scalar equations (43) and (44) cannot describe the motion of a rigid body correctly. Therefore, the governing equations (43) and (44) cannot be used in continuum mechanics to describe the motion of infinitesimal elements of matter.

We also realize that we cannot use higher moments that arbitrarily, and consider for instance the governing equations based on the force and invalid first and second symmetric moment equations

$$\sum F_i^{ext} = \sum ma_i \tag{45}$$

$$\sum \tilde{M}_{ij}^{ext} = \sum \frac{1}{2}\left(ma_i x_j + ma_j x_i\right) \tag{46}$$

$$\sum \tilde{M}_{ijk}^{ext} = \sum \frac{1}{2}\left(ma_i x_j + ma_j x_i\right) x_k$$
$$= \frac{d}{dt}\sum \tilde{\tilde{L}}_{ijk} - \sum \left[\frac{1}{2}\left(mv_i v_j + mv_j v_i\right) x_k + \frac{1}{2}\left(mv_i x_j + mv_j x_i\right) v_k\right] \tag{47}$$

where $\sum \tilde{M}_{ij}^{int}$ and $\sum \tilde{M}_{ijk}^{int}$ are neglected without any justification.

Furthermore, we should notice that there is no symmetry of space corresponding to the conservation of the first general, first symmetric, or second symmetric moment of momentum $\sum \hat{L}_{ij}$, $\sum \tilde{L}_{ij}$ and $\sum \tilde{\tilde{L}}_{ijk}$. This clearly shows that the first and second moment equations (46) and (47) are incorrect. Therefore, the symmetry of space requires that the equations (37) and (38) be the only governing equations for the mechanics of a system of particles.

Higher gradient theories are based on higher moment equations, which resemble the higher non-physical moment equations presented here. Consequently, these theories cannot represent physical reality.

## 4. Fundamental governing equations of motion for a continuum

Here we present a summary and the governing equations of continuum mechanics in the framework of couple stress theory (CST) (Cosserat and Cosserat, 1909; Mindlin and Tiersten, 1962; Koiter, 1964; Hadjesfandiari and Dargush, 2011). This includes the original Mindlin-Tiersten-Koiter couple stress theory (MTK-CST), and consistent couple stress theory (C-CST). We also present these theories by using the third order true tensor form of the couple-stress tensor.

### 4.1. Original couple stress theory

The original MTK-CST is a fundamental pillar in the development of size-dependent continuum mechanics. This theory is an extension of rigid body mechanics, which then is recovered in the absence of deformation (Hadjesfandiari and Dargush, 2015b).

Consider a material continuum occupying a volume $V$ bounded by a surface $S$. The deformation of the body is represented by the continuous displacement field $u_i$. It should be noticed that for the continuum the velocity and acceleration fields are $v_i = \frac{Dx_i}{Dt}$ and $a_i = \frac{Dv_i}{Dt}$, respectively, where $\frac{D}{Dt}$ is the material or substantial derivative. However, in small deformation theory, we can use the approximation $v_i = \frac{\partial u_i}{\partial t} = \dot{u}_i$ and $a_i = \frac{\partial^2 u_i}{\partial t^2} = \ddot{u}_i$.

In infinitesimal deformation theory, the displacement vector field $u_i$ is sufficiently small that the infinitesimal strain and rotation tensors are defined as

$$e_{ij} = u_{(i,j)} = \frac{1}{2}(u_{i,j} + u_{j,i}) \tag{48}$$

$$\omega_{ij} = u_{[i,j]} = \frac{1}{2}(u_{i,j} - u_{j,i}) \tag{49}$$

respectively. Since the true (polar) rotation tensor $\omega_{ij}$ is skew-symmetrical, one can introduce its corresponding dual pseudo (axial) rotation vector as

$$\omega_i = \frac{1}{2}\varepsilon_{ijk}\omega_{kj} = \frac{1}{2}\varepsilon_{ijk}u_{k,j} \tag{50}$$

In consistent continuum mechanics, we consider the rigid body portion of motion of infinitesimal elements of matter (or rigid triads) at each point of the continuum (Hadjesfandiari and Dargush, 2015b). Hence, the governing equations describe the motion of infinitesimal elements of matter at each point. Therefore, the degrees of freedom are the displacements $u_i$ and rotations $\omega_i$ at each point, which describe translation and rotation, respectively, of an infinitesimal element of matter in the neighborhood of the point. However, the continuity of matter within the continuum description restrains the rotation field $\omega_i$ by the relation (50). This of course shows that the rotation field $\omega_i$ is not independent of the displacement field $u_i$.

For the continuum in MTK-CST, the internal stresses are represented by true (polar) force-stress $\sigma_{ij}$ and pseudo (axial) couple-stress $\mu_{ij}$ tensors (Cosserat and Cosserat, 1909), which each have at most nine independent components. The force-traction vector $t_i^{(n)}$ and couple-traction vector $m_i^{(n)}$ through a surface element $dS$ in the volume with outward directed unit normal $n_i$ are given as

$$t_i^{(n)} = \sigma_{ji}n_j \tag{51}$$

$$m_i^{(n)} = \mu_{ji}n_j \tag{52}$$

The fundamental governing equations in couple stress theory (CST) are based on the force and moment equations for system of particles (37) and (38) represented here as

$$\sum F_i^{ext} = \sum ma_i \qquad \sum \mathbf{F}^{ext} = \sum m\mathbf{a} \qquad (53)$$

$$\sum M_i^{ext} = \varepsilon_{ijk} \sum x_j ma_k \qquad \sum \mathbf{M}^{ext} = \sum \mathbf{x} \times m\mathbf{a} \qquad (54)$$

Therefore, we consider an arbitrary part of the material continuum occupying a volume $V_a$ enclosed by boundary surface $S_a$. The force and moment equations for this part of the body are

$$\int_{S_a} \mathbf{t}^{(n)} dS + \int_{V_a} \mathbf{f} dV = \int_{V_a} \rho \mathbf{a} dV \qquad (55)$$

$$\int_{S_a} \left[ \mathbf{x} \times \mathbf{t}^{(n)} + \mathbf{m}^{(n)} \right] dS + \int_{V_a} \mathbf{x} \times \mathbf{f} dV = \int_{V_a} \mathbf{x} \times \rho \mathbf{a} dV \qquad (56)$$

where $f_i$ is the specified body-force density and $\rho$ is the mass density. In terms of components, these equations become

$$\int_{S_a} t_i^{(n)} dS + \int_{V_a} f_i dV = \int_{V_a} \rho a_i dV \qquad (57)$$

$$\int_{S_a} [\varepsilon_{ijk} x_j t_k^{(n)} + m_i^{(n)}] dS + \int_{V_a} \varepsilon_{ijk} x_j f_k \, dV = \int_{V_a} \varepsilon_{ijk} x_j \rho a_k dV \qquad (58)$$

By using the relations (51) and (52) for tractions, along with the divergence theorem, and noticing the arbitrariness of volume $V_a$, we finally obtain the differential form of the governing equations of motion for an infinitesimal element of matter as

$$\sigma_{ji,j} + f_i = \rho a_i \qquad (59)$$

$$\mu_{ji,j} + \varepsilon_{ijk} \sigma_{jk} = 0 \qquad (60)$$

It should be noticed that the derivatives of stresses in the governing equations (59) and (60) are of first order. This is the character of the general fundamental laws of continuum mechanics that their basic form should have first derivatives of stresses, not second or higher orders.

We observe that since $u_i$ and $\omega_i$ are the degrees of freedom, the governing equations describing the translational and rotational motion of element of matter are the vectorial force and vectorial moment equations, where

$$u_i \to 3 \text{ force equations} \qquad \sigma_{ji,j} + f_i = \rho \ddot{u}_i \qquad (61)$$

$$\omega_i \to 3 \text{ moment equations} \qquad \mu_{ji,j} + \varepsilon_{ijk}\sigma_{jk} = 0 \qquad (62)$$

The generally non-symmetric force-stress tensor can be decomposed as

$$\sigma_{ij} = \sigma_{(ij)} + \sigma_{[ij]} \qquad (63)$$

where $\sigma_{(ij)}$ and $\sigma_{[ij]}$ are the symmetric and skew-symmetric parts of this tensor, respectively. We notice that the moment equation (60) gives the skew-symmetric part of the force-stress tensor as

$$\sigma_{[ji]} = \frac{1}{2}\varepsilon_{ijk}\mu_{lk,l} \qquad (64)$$

Thus, for the total force-stress tensor, we have

$$\sigma_{ji} = \sigma_{(ji)} + \frac{1}{2}\varepsilon_{ijk}\mu_{lk,l} \qquad (65)$$

By carrying this relation into the linear governing equation (59), we obtain

$$\left[\sigma_{(ji)} + \frac{1}{2}\varepsilon_{ijk}\mu_{lk,l}\right]_{,j} + f_i = \rho a_i \qquad (66)$$

which can be called the reduced linear governing equation. Since this equation is a combination of the basic force and moment equations (59) and (60), it cannot be considered as a fundamental law by itself. This can be confirmed by noticing that the highest derivative in the governing equation (66) is of second order.

Mindlin and Tiersten (1962) and Koiter (1964) have shown that the displacement field $u_i$ prescribed on a smooth part of the boundary $S$, specifies the normal component of the rotation $\omega^{(nn)} = \omega_i n_i$. Accordingly, they have demonstrated that material in a consistent couple stress theory does not support independent distributions of normal surface couple-traction $m^{(nn)} = m_i^{(n)} n_i$. This means

$$m^{(nn)} = m_i^{(n)} n_i = \mu_{ji} n_i n_j = 0 \tag{67}$$

Consequently, Mindlin and Tiersten (1962) and Koiter (1964) correctly established that five geometrical and five mechanical boundary conditions could be specified on a smooth surface. However, they did not realize the fundamental implication of (67) on the character of the couple-stress tensor and its energetically conjugate curvature tensor. Unfortunately, Mindlin and Tiersten (1962) and Koiter (1964) confused the matter by introducing the approximate reduced boundary condition method from structural mechanics (Hadjesfandiari and Dargush, 2011, 2015b). As a result, in their development, the form of the couple-stress tensor $\mu_{ij}$ is arbitrary and the gradient of the rotation

$$k_{ij} = \omega_{j,i} \tag{68}$$

is considered as the curvature tensor measure of deformation. Since the tensor $k_{ij}$ describes the combination of bending and twisting of elements of the continuum, it may be called the bend–twist tensor (deWit, 1970).

The original Mindlin-Tiersten-Koiter (MTK-CST) theory suffers from some serious inconsistencies and difficulties with the underlying formulations (Eringen, 1968; Hadjesfandiari and Dargush, 2011, 2015a,b, 2016), which may be summarized as follows:

1. The inconsistency in boundary conditions, since the normal component of the couple-traction vector $m^{(nn)} = m_i n_i$ appears in the formulation violating the condition (67);

2. The appearance of an indeterminate spherical part in the couple-stress tensor, and thus, in the skew-symmetric part of the force-stress tensor;

3. The disturbing appearance of the body couple in the constitutive relation for the force-stress tensor (Hadjesfandiari and Dargush, 2011, 2015a);

4. The appearance of too many constitutive coefficients, which makes the MTK-CST less attractive from a practical perspective. It turns out there are 105 couple-stress elastic coefficients for linear elastic anisotropic material (Mindlin and Tiersten, 1962; Hadjesfandiari and Dargush, 2011, 2015a). For linear elastic isotropic material, these reduce to two couple-stress elastic coefficients $\eta$ and $\eta'$, although only one of these elastic coefficients, $\eta$, appears in the final governing equations when written in terms of displacements.

Eringen (1968) realized the indeterminacy problem as a major mathematical inconsistency in the original Mindlin-Tiersten-Koiter couple stress theory (MTK-CST), which he afterwards called indeterminate couple stress theory.

*4.2. Original couple stress theory based on true tensors*

As explained, the fundamental governing equations for a system of particles can also be written in the form of force and skew-symmetric moment equations

$$\sum F_i^{ext} = \sum ma_i \tag{69}$$

$$\sum M_{ij}^{ext} = \sum \frac{1}{2}\left(ma_i x_j - ma_j x_i\right) \tag{70}$$

It will be very instructive if we present all the quantities and governing equations in couple stress theory (CST) based on this form of fundamental laws in terms of true tensors. This presentation of MTK-CST is analogous to other higher gradient theories, such as distortion gradient and strain gradient theories and, thus, will permit more straightforward comparisons.

In this presentation, the interaction is specified by means of the force vector $t_i^{(n)}dS$ and a skew-symmetric moment tensor $m_{ij}^{(n)}dS$. The tensor $m_{ij}^{(n)}$ is the skew-symmetric true couple-traction tensor, where

$$m_{ji}^{(n)} = -m_{ij}^{(n)} \tag{71}$$

We notice the dual relations between the pseudo (axial) vector $m_i^{(n)}$ and the true (polar) tensor $m_{ij}^{(n)}$ are

$$m_{ij}^{(n)} = \frac{1}{2}\varepsilon_{jik}m_k^{(n)}, \qquad m_i^{(n)} = \varepsilon_{ijk}m_{kj}^{(n)} \tag{72}$$

The internal stresses are then represented by generally non-symmetric true (polar) force-stress $\sigma_{ij}$ and true couple-stress $\mu_{ijk}$ tensors, where

$$\mu_{ikj} = -\mu_{ijk} \tag{73}$$

The third order true tensor $\mu_{ijk}$ is related to the second order pseudo (axial) tensor $\mu_{ij}$ with the dual relations

$$\mu_{ijk} = \frac{1}{2}\varepsilon_{kjl}\mu_{il}, \qquad \mu_{ij} = \varepsilon_{lkj}\mu_{ikl} \tag{74}$$

Consequently, the force-traction vector $t_i^{(n)}$ and couple-traction vector $m_i^{(n)}$ can be written as

$$t_i^{(n)} = \sigma_{ji}n_j \tag{75}$$

$$m_{ij}^{(n)} = \mu_{kij}n_k \tag{76}$$

In MTK-CST, the tensors $m_{ij}^{(n)}$ and $\mu_{ijk}$ are specified by three and nine components, respectively.

We notice that the moment equation (60) gives

$$\sigma_{[ji]} = \frac{1}{2} \varepsilon_{ijk} \mu_{lk,l} = \frac{1}{2} \varepsilon_{ijk} \varepsilon_{nmk} \mu_{lmn,l}$$
$$= \frac{1}{2}\left(\delta_{in}\delta_{jm} - \delta_{im}\delta_{jn}\right)\mu_{lmn,l} = \frac{1}{2}\left(\mu_{lji,l} - \mu_{lij,l}\right) \quad (77)$$
$$= -\mu_{lij,l}$$

Therefore, the governing equations for an infinitesimal element of matter become

$$\sigma_{ji,j} + f_i = \rho a_i \quad (78)$$

$$\sigma_{[ji]} + \mu_{kij,k} = 0 \quad (79)$$

Notice that the moment equation (79) still represents only three scalar equations. Interestingly, by combining the governing equations (78) and (79), we obtain the second order reduced linear governing equation of motion as

$$\left[\sigma_{(ij)} - \mu_{kij,k}\right]_{,j} + f_i = \rho a_i \quad (80)$$

It seems more enlightening if we derive the governing moment equation (79) from scratch. For this purpose, we consider the skew-symmetric moment equation (70) for the material in the arbitrary volume $V_a$ enclosed by the boundary surface $S_a$ as

$$\int_{S_a} \left[\frac{1}{2}\left(x_j t_i^{(n)} - x_i t_j^{(n)}\right) + m_{ij}^{(n)}\right] dS + \int_{V_a} \frac{1}{2}\left(x_j f_i - x_i f_j\right) dV = \int_{V_a} \frac{1}{2}\left(x_j \rho a_i - x_i \rho a_j\right) dV \quad (81)$$

By using the expressions (75) and (76) for tractions, this becomes

$$\int_{S_a} \left[\frac{1}{2}\left(x_j \sigma_{ki} n_k - x_i \sigma_{kj} n_k\right) + \mu_{kij} n_k\right] dS + \int_{V_a} \frac{1}{2}\left(x_j f_i - x_i f_j\right) dV$$
$$= \int_{V_a} \frac{1}{2}\left(x_j \rho a_i - x_i \rho a_j\right) dV \quad (82)$$

Using the divergence theorem and some manipulation, we obtain

$$\int_{V_a} \left[\frac{1}{2}\left(\delta_{jk}\sigma_{ki} + x_j \sigma_{ki,k} - \delta_{ik}\sigma_{kj} - x_i \sigma_{kj,k}\right) + \mu_{kij,k}\right] dV + \int_{V_a} \frac{1}{2}\left(x_j f_i - x_i f_j\right) dV$$
$$= \int_{V_a} \frac{1}{2}\left(x_j \rho a_i - x_i \rho a_j\right) dV \quad (83)$$

or

$$\int_{V_a}\left[\frac{1}{2}(\sigma_{ji}-\sigma_{ij})+\frac{1}{2}x_j(\sigma_{ki,k}+f_i-\rho a_i)-\frac{1}{2}x_i(\sigma_{kj,k}+f_j-\rho a_j)+\mu_{kij,k}\right]dV=0 \quad (84)$$

Using the force governing equation (78), this becomes

$$\int_{V_a}\left[\frac{1}{2}(\sigma_{ji}-\sigma_{ij})+\mu_{kij,k}\right]dV=0 \quad (85)$$

By noticing the arbitrariness of volume $V_a$, we obtain

$$\sigma_{[ji]}+\mu_{kij,k}=0 \quad (86)$$

which is exactly the final skew-symmetric moment equation (79).

Interestingly, the principle of virtual work in MTK-CST becomes

$$\int_S\left[t_i^{(n)}\delta u_i+m_{ij}^{(n)}\delta\omega_{ij}\right]dS+\int_V f_i\delta u_i dV=\int_V\left[\sigma_{ji}\delta e_{ij}+\mu_{kij}\delta\omega_{ij,k}\right]dV+\int_V\rho a_i\,\delta u_i\,dV \quad (87)$$

This relation shows that the third order tensor bend-twist tensor $k_{ijk}$ can be defined as

$$k_{ijk}=\omega_{ij,k} \quad (88)$$

with the symmetry relation

$$k_{jik}=-k_{ijk} \quad (89)$$

Interestingly, we have the additional constraint relations

$$k_{iik}=0, \qquad k_{ijk}+k_{jki}+k_{kij}=0 \quad (90)$$

The dual relations between the second and third order bend-twist tensors are

$$k_{ijk}=\varepsilon_{jip}k_{kp}, \qquad k_{ij}=\frac{1}{2}\varepsilon_{jpq}k_{qpi} \quad (91)$$

Consequently, the principle of virtual work (87) for MTK-CST can be written as

$$\int_S\left[t_i^{(n)}\delta u_i+m_{ij}^{(n)}\delta\omega_{ij}\right]dS+\int_V f_i\delta u_i dV=\int_V\left[\sigma_{ji}\delta e_{ij}+\mu_{kij}\delta k_{ijk}\right]dV+\int_V\rho a_i\,\delta u_i\,dV \quad (92)$$

## 5. Consistent couple stress theory

After almost half a century of confusion created by the indeterminacy of MTK-CST, Hadjesfandiari and Dargush (2011, 2015b) discovered consistent couple stress theory (C-CST). This theory not only answers the criticism of Eringen about the indeterminacy, but also resolves other inconsistencies in the original MTK-CST. The main achievement of this development is discovering the subtle skew-symmetric character of the couple-stress tensor

$$\mu_{ji} = -\mu_{ij} \qquad (93)$$

The fundamental step in this development is satisfying the requirement (67) that the normal component of the couple-traction vector must vanish on the boundary surface in a systematic way, i.e. $m^{(nn)} = \mu_{ji} n_j n_i = 0$. This is what Mindlin, Tiersten and Koiter missed in their important developments of MTK-CST, although they correctly established the consistent boundary conditions. They did not realize that their consistent boundary conditions simply shows the existence of the normal twisting couple-traction $m^{(nn)} = \mu_{ji} n_j n_i$ is physically impossible.

We notice that in C-CST the skew-symmetric part of the bend-twist tensor

$$\kappa_{ij} = \omega_{[i,j]} = \frac{1}{2}\left(\omega_{i,j} - \omega_{j,i}\right) \qquad (94)$$

is the curvature tensor measure of deformation. This tensor, which is called the mean curvature tensor, is energetically conjugate to the skew-symmetric couple-stress tensor $\mu_{ij}$. It turns out that the skew-symmetric tensors $\mu_{ij}$ and $\kappa_{ij}$ can be represented by their dual true couple-stress vector $\mu_i$ and true mean curvature vector $\kappa_i$ (Hadjesfandiari and Dargush, 2011), where

$$\mu_i = \frac{1}{2}\varepsilon_{ijk}\mu_{kj} \qquad (95)$$

$$\kappa_i = \frac{1}{2}\varepsilon_{ijk}\kappa_{kj} \tag{96}$$

Therefore, the consistent skew-symmetric couple stress theory (C-CST) may be called the vector couple stress theory. Interestingly, the mean curvature vector $\kappa_i$ can also be expressed as

$$\kappa_i = \frac{1}{2}\omega_{ji,j} = \frac{1}{4}\left(u_{j,ji} - \nabla^2 u_i\right) = \frac{1}{2}\left(e_{jj,i} - e_{ij,j}\right) \tag{97}$$

It is astounding to note that the skew-symmetric character of the couple-stress tensor immediately resolves the indeterminacy problem by establishing that there is no spherical component. As a result, the couple-stress tensor is determinate in the skew-symmetric C-CST. Interestingly, there is an interrelationship between the consistent mechanical boundary condition (67), $m^{(nn)} = 0$, and the determinacy of the couple-stress tensor; resolving one, resolves the other (Hadjesfandiari and Dargush, 2011, 2015a,b).

The governing equations (59) and (60) for an infinitesimal element of matter can be written as

$$\sigma_{ji,j} + f_i = \rho a_i \tag{98}$$

$$\sigma_{[ji]} + \mu_{[i,j]} = 0 \tag{99}$$

The moment equations (99) give the skew-symmetric part of the force-stress tensor. Thus, for the total force-stress tensor, we have

$$\sigma_{ji} = \sigma_{(ji)} - \mu_{[i,j]} \tag{100}$$

As a result, the reduced linear governing equation becomes

$$\left[\sigma_{(ji)} + \mu_{[j,i]}\right]_{,j} + f_i = \rho \ddot{u}_i \tag{101}$$

Meanwhile, the surface couple-traction vector $m_i$ can be written as

$$m_i^{(n)} = \mu_{ji}n_j = \varepsilon_{ijk}n_j\mu_k \tag{102}$$

Since this traction is tangent to the surface, it creates bending deformation only.

Interestingly, the virtual work theorem for this formulation is (Hadjesfandiari and Dargush, 2011,2015b)

$$\delta W_{ext} = \delta W_{int} + \int_V \rho a_i \, \delta u_i \, dV \qquad (103)$$

where the external and internal virtual work, respectively, are written

$$\delta W_{ext} = \int_S \left[ t_i^{(n)} \, \delta u_i + m_i^{(n)} \, \delta \omega_i \right] dS + \int_V f_i \, \delta u_i \, dV \qquad (104)$$

$$\delta W_{int} = \int_V \left[ \sigma_{(ji)} \, \delta e_{ij} - 2\mu_i \, \delta \kappa_i \right] dV \qquad (105)$$

Therefore, the virtual work theorem for this formulation becomes

$$\int_S \left[ t_i^{(n)} \, \delta u_i + m_i^{(n)} \, \delta \omega_i \right] dS + \int_V f_i \, \delta u_i \, dV$$
$$= \int_V \left[ \sigma_{(ji)} \, \delta e_{ij} - 2\mu_i \, \delta \kappa_i \right] dV + \int_V \rho a_i \delta u_i dV \qquad (106)$$

In C-CST, we also have the dual relations between the true couple-stress vector $\mu_i$ and a third order true couple-stress tensor $\mu_{ijk}$ as

$$\mu_{ijk} = \frac{1}{2}\left(\delta_{ik}\mu_j - \delta_{ij}\mu_k\right), \qquad \mu_i = -\mu_{kki} \qquad (107)$$

These relations show that in C-CST we have

$$\mu_{ijk} + \mu_{jki} + \mu_{kij} = 0 \qquad (108)$$

Interestingly, it is also noticed

$$\mu_{ji}\delta\kappa_{ij} = -\mu_{kij}\varepsilon_{lij}\delta\kappa_{kl} \qquad (109)$$

This shows that the third order mean curvature tensor $\kappa_{ijk}$ can be defined as

$$\kappa_{ijk} = \varepsilon_{jil}\kappa_{kl} \qquad (110)$$

with the additional properties

$$\kappa_{iik} = 0, \qquad \kappa_{jik} = -\kappa_{ijk} \qquad (111)$$

We notice the dual relation for $\kappa_{ij}$ and $\kappa_{ijk}$ is

$$\kappa_{ij} = \frac{1}{2}\varepsilon_{kli}\kappa_{klj} \tag{112}$$

One also can obtain the dual relations

$$\kappa_{ijk} = \delta_{jk}\kappa_i - \delta_{ik}\kappa_j, \qquad \kappa_i = \frac{1}{2}\kappa_{ikk} \tag{113}$$

Therefore, the third order mean curvature tensor $\kappa_{ijk}$ in terms of the second order rotation tensor $\omega_{ij}$ and the third order bend–twist tensor $k_{ijk}$, respectively, becomes

$$\begin{aligned}\kappa_{ijk} &= \frac{1}{2}\left(\delta_{jk}\omega_{pi,p} - \delta_{ik}\omega_{pj,p}\right) \\ &= \frac{1}{2}\left(\delta_{jk}\delta_{iq} - \delta_{ik}\delta_{jq}\right)\omega_{pq,p}\end{aligned} \tag{114}$$

and

$$\kappa_{ijk} = \frac{1}{2}\left(\delta_{jk}\delta_{iq} - \delta_{ik}\delta_{jq}\right)k_{pqp} \tag{115}$$

We notice that (109) can be written as

$$-2\mu_i\delta\kappa_i = \mu_{ji}\delta\kappa_{ij} = \mu_{kji}\delta\kappa_{ijk} \tag{116}$$

Therefore, the principle of virtual work in C-CST becomes

$$\int_S \left[t_i^{(n)}\delta u_i + m_{ij}^{(n)}\delta\omega_{ij}\right]dS + \int_V f_i\delta u_i dV = \int_V \left[\sigma_{ji}\delta e_{ij} + \mu_{kji}\delta\kappa_{ijk}\right]dV + \int_V \rho a_i\,\delta u_i\,dV \tag{117}$$

As can be noticed, the virtual work theorem (117) requires that the measures of deformation be $e_{ij}$ and $\kappa_{ijk}$, corresponding to $\sigma_{ij}$ and $\mu_{ijk}$, respectively, that is

$$e_{ij} \leftrightarrow \sigma_{ji} \tag{118}$$

$$\kappa_{ijk} \leftrightarrow \mu_{kji} \tag{119}$$

The presentation of C-CST by true third order couple-stress $\mu_{ijk}$ and mean curvature $\kappa_{ijk}$ tensors makes the comparison of C-CST with other higher gradient theories, such as distortion gradient and strain gradient theories, more straightforward. This will be pursued in a subsequent paper.

It is important to note that C-CST answers the criticism of Eringen about the indeterminacy of the couple-stress tensor in MTK-CST without adding any new artificial law. Thus, C-CST systematically links efforts of the Cosserats, Mindlin, Tiersten and Koiter and others over a span of a century.

The following observations can be made from the discovery of C-CST, which demonstrate the inner beauty and natural simplicity of consistent continuum mechanics:

1. In classical continuum mechanics, there are no couple-stresses, such that $\mu_{ij} = 0$. As a result, the force-stress tensor $\sigma_{ij}$ is symmetric.

2. In size-dependent continuum mechanics, the force-stress tensor $\sigma_{ij}$ is not symmetric, whereas the couple-stress tensor $\mu_{ij}$ is skew-symmetric.

This result shows that both classical and size-dependent continuum mechanics enjoy some level of symmetry in their inner structures. Hadjesfandiari et al. (2015) provides the theoretical background of C-CST for any continuum, including both solids and fluids. Therefore, C-CST offers a fundamental basis for the development of size-dependent theories in many multi-physics disciplines that may govern the behavior of continua at the smallest scales. For instance, Hadjesfandiari (2013, 2014) has developed size-dependent piezoelectricity and thermoelasticity. Remarkably, C-CST has recently demonstrated its self-inconsistency by resolving all issues in the existing continuous defect theory, which has resulted in consistent continuous defect theory (Hadjesfandiari and Dargush, 2018). In this theory, the dislocation density tensor is skew-symmetric and can be represented by a vector. This development also establishes the dualism between geometry and statics of consistent continuous defect theory based on C-CST.

## 6. Conclusions

In this paper, we have examined the physical and mathematical consistencies of the governing equations in continuum mechanics. We have demonstrated that based on Newton's third law of action and reaction, there are only two vectorial governing equations of motion for a system of particles and a continuum. These are the force and skew-symmetric moment governing equations, which provide the governing equations of motion for infinitesimal elements of matter at each point. The deformation in consistent continuum mechanics is completely specified by the continuous displacement field $u_i$. Therefore, the displacement field $u_i$ is the translation degrees of freedom at each point, and half of its curl, the vector $\omega_i$ is the rotation degrees of freedom. Higher moment governing equations are non-physical violating Newton's third law of action and reaction, and the angular momentum theorem. Furthermore, this shows that higher gradient theories, such as distortion gradient and strain gradient theories, based on these non-physical moment equations also are non-physical. This will be demonstrated in detail in a future work.

On the other hand, consistent couple stress theory (C-CST) is based on the consistent set of governing equations, incorporating the force and (skew-symmetric) moment governing equations. The triumph of this theory is establishing the skew-symmetric character of the couple-stress tensor. This theory is grounded implicitly on the rigid body portion of motion of infinitesimal elements of matter at each point of the continuum. Therefore, C-CST is an extension of rigid body mechanics, which then is recovered in the absence of deformation. Thus, the determinate consistent couple stress theory provides a fundamental basis for the development of many linear and non-linear size-dependent multi-physics phenomena in continuum mechanics.